\documentclass[reprint]{revtex4-1}
\usepackage{gensymb,graphicx}
\usepackage{units}
\begin{document}
\title{Driving field amplitude gauged quantitative inverse spin Hall effect detection}
\author{M. Kavand}
\affiliation{Department of Physics and Astronomy, University of Utah}
\author{C. Zhang}
\affiliation{Department of Physics and Astronomy, University of Utah}
\author{D. Sun}
\affiliation{Department of Physics and Astronomy, University of Utah}
\author{H. Malissa}
\email{hans.malissa@utah.edu}
\affiliation{Department of Physics and Astronomy, University of Utah}
\author{Z. V. Vardeny}
\affiliation{Department of Physics and Astronomy, University of Utah}
\author{C. Boehme}
\email{boehme@physics.utah.edu}
\affiliation{Department of Physics and Astronomy, University of Utah}
\date{\today}
\begin{abstract}
Spin transport in thin-film materials can be studied by ferromagnetic resonantly (FMR) driven spin pumping of a charge-free spin current which induces an electromotive force through the inverse spin Hall effect (ISHE). For quantitative ISHE experiments, precise control of the FMR driving field amplitude $B_1$ is crucial. This study exploits in situ monitoring of $B_1$ by utilization of electron paramagnetic resonantly (EPR) induced transient nutation of paramagnetic molecules (a 1:1 complex of $\alpha$,$\gamma$-bisdiphenylene-$\beta$-phenylallyl and benzene, BDPA) placed as $B_1$ probe in proximity of a NiFe/Pt-based ISHE device. Concurrent to an ISHE experiment, $B_1$ is obtained from the inductively measured BDPA Rabi-nutation frequency. Higher reproducibility is achieved by renormalization of the ISHE voltage to $B_1^2$ with an accuracy that is determined by the homogeneity of the FMR driving field and thus by the applied microwave resonator and ISHE device setup.
\end{abstract}
\maketitle

The detection of spin pumping through the inverse spin Hall effect (ISHE) in semiconducting materials under ferromagnetic resonant (FMR) excitation of an adjacent ferromagnet by microwave (MW) radiation is a recently developed technique for the study of spin-transport phenomena and spin-orbit coupling (SOC) effects in various materials \cite{dali_pero,saitoh_nat_commun,dali_osec,sirringhaus_nat_mater_2013,saitoh_JApplPhys}, in particular in materials with weak SOC, such as organic semiconductors. As the magnitude of the ISHE response scales linearly with the strength of the MW radiation field \cite{dali_osec,sirringhaus_nat_phys_2014,sirringhaus_nat_mater_2013,saitoh_JApplPhys},  \emph{quantitative} ISHE experiments which, for instance, are needed to test fundamental theories of charge free spin-injection into organic semiconductors \cite{raikh} or the nature of spin transport in the latter \cite{yu}, require a precise knowledge of the strength of the MW radiation amplitude $B_1$. Most studies involving ISHE experiments derive $B_1$ from estimates that are based on the MW power applied to the resonator or waveguide structure. However, this approach is generally not reliable because the conversion factor that translates a MW power applied to a resonator to $B_1^2$ depends on the particular resonator and device geometries and also, it depends on how a device is placed within a resonator. Smallest position changes can cause significant changes of $B_1^2$, even if nominally identical conditions exist.

In this work, we study the distribution of $B_1$ values for a dielectric cylindrical microwave resonator and present as well as evaluate a method for the accurate determination of $B_1$ very close to the position of an ISHE device. The idea is to use electron paramagnetic resonantly (EPR) induced transient nutation of a spin standard that is mounted directly on the ISHE device, in close proximity of the active region where spin-pumping takes place. The measurement of transient nutation is a well established pulsed EPR technique for the determination \cite{schweiger_jeschke} of the magnitude of $B_1$.

\begin{figure}
\includegraphics[width=\columnwidth]{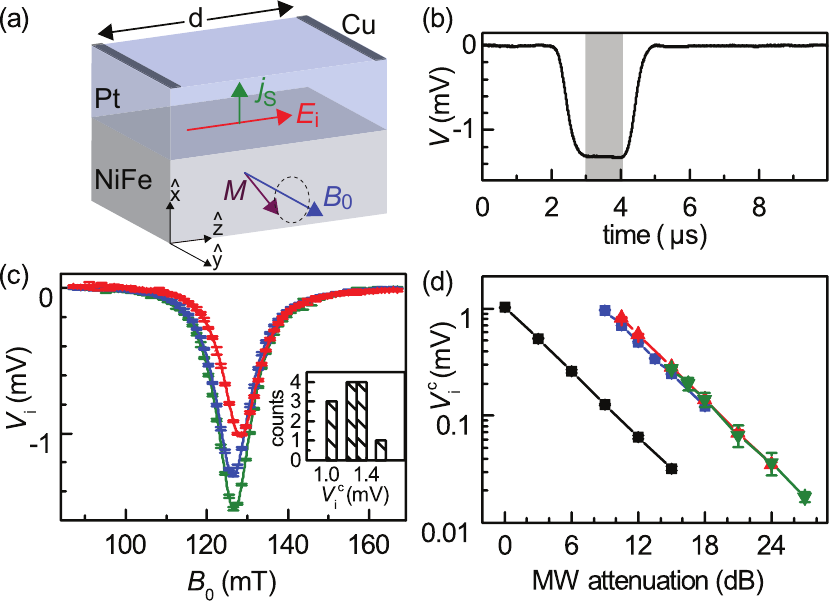}
\caption{\label{fig1} (a) Illustration of the NiFe/Pt/Cu device used for the ISHE experiments. $\mathbf{B}_0$ is the external magnetic field, $\mathbf{M}$ is the dynamic magnetization due to FMR, $\mathbf{j}_\mathrm{S}$ is the spin current, and $\mathbf{E}_\mathrm{i}$ the electric field due to the ISHE. (b) Measurement of the ISHE Voltage $V_\mathrm{i}$ as function of time during the pulsed MW excitation. The shaded region indicates the time interval during the \unit[2]{$\mu$s} pulse over which the transient was averaged. (c) $V_\mathrm{i}$ measured multiple times as functions of $B_0$ on one and the same device under nominally identical conditions (3 of 12 measured traces are shown). The symbols represent data points acquired in the way illustrated in (b), the error bars correspond to the standard deviation of the average over the transient measurements. The solid lines represent Lorentzian fits of the FMR driven ISHE voltage $V_\mathrm{i}$. Inset: histogram of the ISHE voltage values $V_\mathrm{i}^\mathrm{c}$ measured on resonance. (d) Measurements of $V_\mathrm{i}^\mathrm{c}$ for different resonator coupling adjustments (black: $Q\approx 200$, blue: $Q\approx 400$, green: $Q\approx 800$, red: $Q\approx 1000$) and at different MW powers. The horizontal axis shows the attenuation setting of the pulsed \unit[1]{kW} MW source.}
\end{figure}

\begin{figure}
\includegraphics[width=\columnwidth]{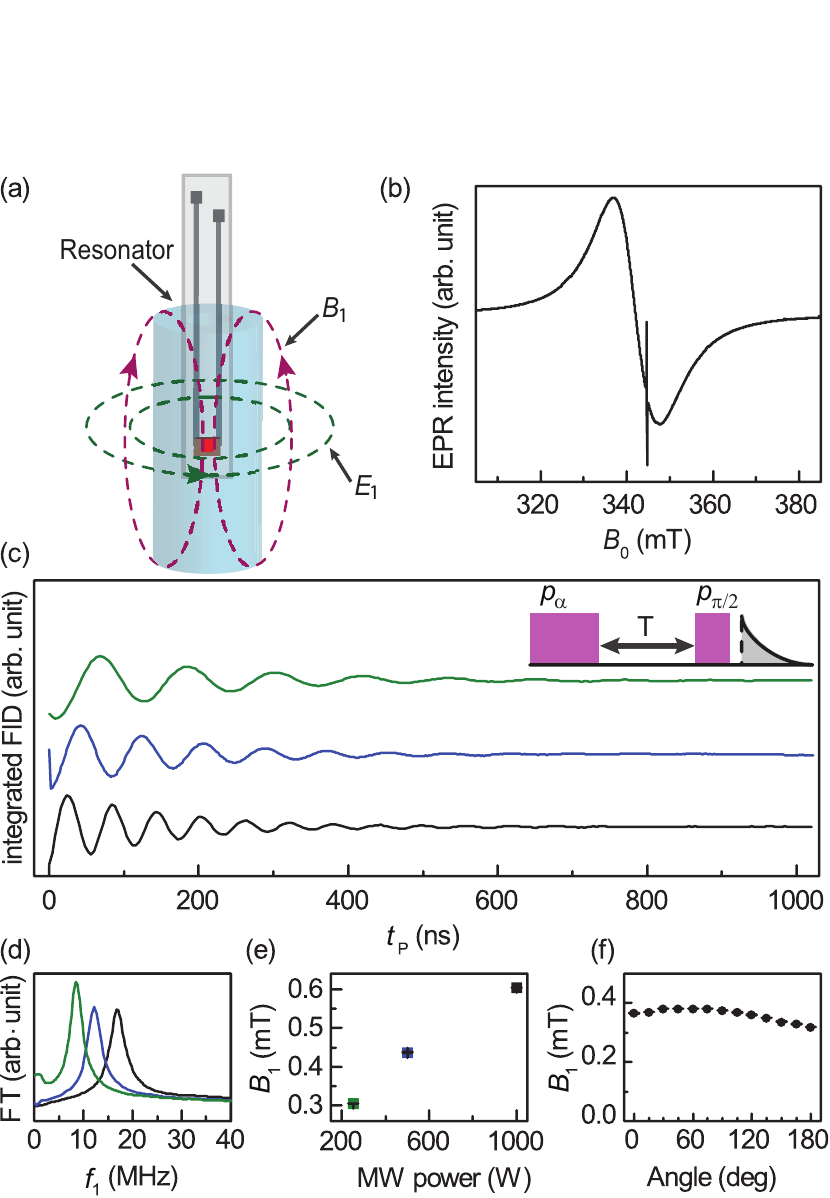}
\caption{\label{fig2} (a) Sketch of an ISHE device with the BDPA grain (red) attached at the center of the cylindrical dielectric MW resonator, for which the mode pattern of $B_1$ and $E_1$ are illustrated by the red and green dashed lines. (b) Magnetic field modulated lock-in detected MW absorption spectrum of the ISHE device with an attached BDPA grain, recorded for a MW power of \unit[19.7]{mW} with a device surface normal orientation of 85\degree with respect to $B_0$. Both, the FMR signal of the NiFe magnet (broad line) and the EPR signal of the BDPA standard (narrow line) are visible. (c) Transient nutation experiments on BPDA detected by FID measurements at several attenuator settings. The inset shows the pulse sequence that was used to measure the transient nutation. (d) Plots of the absolute values of the Fourier transforms of the transient nutation data. (e) The measured $B_1$ as a function of MW power. (f) The measured values of $B_1$ as functions of the sample surface normal orientations with regard to $\mathbf{B}_0$.}
\end{figure}

The experimental setup for spin pumping and ISHE detection used in this study are shown in Fig.~\ref{fig1}(a) and are described in detail elsewhere \cite{dali_osec}. The ISHE device consists of a ferromagnetic layer (NiFe, \unit[15]{nm}) adjacent to a nonmagnetic layer (Pt, \unit[7]{nm}) that is contacted by two electrodes with a distance $d=\unit[1]{mm}$ [cf.\ Fig.~\ref{fig1}(a) and Fig.~\ref{fig2}(a)]. The ISHE device is placed in a cylindrical dielectric resonator (Bruker FlexLine ER 4118 X-MD5) that is capable of operating under continuous wave (CW) and pulse operation. The $B_1$ to square-root of MW power conversion factor of this resonator is adjustable and it is specified to range from \unit[0.42]{mT W$^{-1/2}$} for CW operation (high quality factor $Q$) to \unit[0.10]{mT W$^{-1/2}$} for pulse operation (low $Q$). The resonator is placed in the external magnetic field $\mathbf{B}_0$ of a Helmholtz-magnet that is part of a commercial X-band pulse EPR spectrometer (Bruker E580). When FMR is established in the NiFe layer, a spin current $\mathbf{j}_\mathrm{S}$ is injected into the nonmagnetic layer, in a direction perpendicular to the interface and $\mathbf{B}_0$. Due to the SOC of Pt this will then induce the ISHE and lead to a lateral electric field $\mathbf{E}_\mathrm{i}$ that is perpendicular to both $\mathbf{j}_\mathrm{S}$ and $\mathbf{B}_0$ and an associated accumulation of charge at the two contacts of the device.

For the given experiment, the magnetic field amplitude $B_1$ of the MW driving field gives rise to the electric field $E_\mathrm{i}=V_\mathrm{i}/d\propto B_1^2$, where $d$ is the distance between the contacts across which the ISHE voltage $V_\mathrm{i}$ is detected \cite{saitoh_JApplPhys_2010,spin_transport}. We can therefore define a conversion factor $\xi_\mathrm{i}=E_\mathrm{i}/B_1^2$ that will be invariant and, in fact, characteristic for a particular ferromagnet to non-ferromagnet interface, as long as thickness effects of the two layers are negligible (i.e. when the thickness of the non-ferromagnet is larger than the spin-diffusion lengths). The conversion factor $\xi_\mathrm{i}$ depends on a number of parameters (see Ref.~\onlinecite{saitoh_JApplPhys_2010}) that are invariant for an interface, in particular the Gilbert damping constant $\alpha$ that depends strongly on the deposition conditions of the ferromagnetic layer, and the saturation magnetization $M_\mathrm{S}$ which is much lower than the $B_0$ at which the experiments were conducted. By independently and accurately measuring both the ISHE voltage $V_\mathrm{i}$ and the driving field amplitude $B_1$, we can establish $\xi_\mathrm{i}$ for any given interface. In experiments that involve inhomogeneous distributions of $B_1$, e.\ g.\ when the MW is delivered by planar waveguide structures \cite{hans_cpw,gajadhar_cpw}, the distribution of $\xi_\mathrm{i}$ will be broader and for quantitative ISHE experiments, such $B_1$ inhomogeneities will pose limitations on the interpretability of the measured voltage.

The need for an independent control of $B_1$ becomes obvious from the data shown in Fig.~\ref{fig1}. Panel (b) shows a transient of $V(t)$ recorded during a \unit[2]{$\mu$s} pulsed MW excitation with a power of nominally \unit[1]{kW}. $V_\mathrm{i}$ is acquired by averaging over the indicated time interval from the transient current response of the device that is measured with a current amplifier (Stanford Research Systems SR570) along with the DC resistivity of the device ($R=\unit[775]{\Omega}$). In Fig.~\ref{fig1}(c), the results of several repetitions of measurements of $V_\mathrm{i}(B_0)$ are displayed. The data was obtained on the same device under nominally identical conditions (temperature, sample orientation, applied MW power, etc.). In between repetitions the sample is removed from the MW resonator and then reinserted into a nominally identical position. The resonator is then retuned in order to establish nominally identical coupling conditions. From the spectra in Fig.~\ref{fig1}(c) it is obvious that there is a substantial amount of variation in the ISHE response between the three displayed data sets. We repeated this procedure 12 times and the resulting peak ISHE voltages $V_\mathrm{i}^\mathrm{c}=V_\mathrm{i}(B_0^\mathrm{c})$ obtained from the measured resonance centers are plotted in the inset of Fig.~\ref{fig1}(c) (with $B_0^\mathrm{c}$ being the value of $B_0$ where the maximum of the ISHE response is detected). For the determination of the resonance centers, the function $V_\mathrm{i}(B_0)$ is fitted with a superposition of a symmetric (ISHE) and antisymmetric (anomalous Hall effect) contribution \cite{dali_osec}. For the random distribution shown in the inset of Fig.~\ref{fig1}(c), we have obtained a relative standard deviation of the measured ISHE of 12.2\%.

In order to test whether the variation of the ISHE voltage is solely caused by undetected changes of the resonator coupling (i.e.\ a change of $Q$ that is due to subtle changes of the sample position but not an intentional change of the resonator coupling adjustment), we repeated measurements of $V_\mathrm{i}^\mathrm{c}$ for several coupling adjustment settings, yet without moving the sample within the resonator, producing different $Q$ values of 1000, 800, 400, and 200, respectively, as obtained from the reflected MW power detected by the EPR spectrometer. Figure~\ref{fig1}(d) shows the results of these measurements in a plot displaying $V_\mathrm{i}^\mathrm{c}$ as a function of MW power (i.e. attenuation from \unit[1]{kW} peak pulse power) for the four different settings of the resonator coupling adjustment. While this data reveals a weak dependence of the ISHE response on $Q$ for $Q\geq 400$, it shows a strongly different response at $Q\approx200$. These results confirm qualitatively similar results previously described in Ref.~\onlinecite{dali_osec}. We conclude from this observation that, while the value of $Q$ is important for the conversion of the applied MW power to $B_1$, in particular for experiments that require low $Q$ (pulsed operation), the influence of small changes of the sample position within the MW resonator as revealed by the data in Fig.~\ref{fig1}(c) can play an even bigger role and consequently, in situ monitoring of $B_1$ becomes necessary.

In order to probe the magnitude of $B_1$, a small volume ($<$\unit[0.1]{$\mu$l}) of the molecular paramagnetic EPR standard BDPA (a crystalline 1:1 complex of $\alpha$,$\gamma$-bisdiphenylene-$\beta$-phenylallyl and benzene \cite{eaton_bdpa_2014,eaton_bdpa_2011}, Sigma-Aldrich, 152560) was mounted directly on top of the NiFe layer of the ISHE device, in close proximity to the active Pt layer. The device was then placed near the center of the resonator [cf.\ Fig.~\ref{fig2}(a)]. Fig.~\ref{fig2}(b) shows a CW electron magnetic resonance spectrum of the device including the BDPA standard sample. Due to the strong anisotropy of the FMR resonance line \cite{dali_osec}, the device orientation with respect to $B_0$ was chosen such that both, the FMR of the NiFe layer and the EPR line of the BDPA standard appear at $B_0$ values that are close to each other.

Free induction decay (FID) detected transient nutation experiment \cite{sherwin_fel_bdpa,astashkin_rabi,mclauchlan_rabi,torrey_rabi,rabi} were conducted on the BDPA standard in order to determine the exact strength of $B_1$ at the position of the device inside the resonator for any given MW power. The MW pulse sequence is illustrated in the inset of Fig.~\ref{fig2}(c): a pulse $\mathrm{p}_\alpha$ with variable duration $t_\mathrm{P}$ nutates the spins on resonance by a tipping angle $\alpha=\omega_1t_\mathrm{P}$ with $\omega_1=2\pi f_1=g\mu_\mathrm{B} B_1/\hbar$ being the Rabi frequency that scales with the strength of $B_1$ ($g$ is the g-factor and $\mu_\mathrm{B}$ the Bohr magneton). After a fixed delay $T=\unit[20]{ns}$, a $\pi/2$ pulse $\mathrm{p}_{\pi/2}$ with a fixed duration is used to induce a FID that is measured by integration of the transient EPR response over \unit[400]{ns}. The intensity of the FID is directly proportional to the magnetization along $B_0$ and therefore $\alpha$. In practice, the duration of the second pulse was kept constant at \unit[16]{ns} for all applied MW powers; a fraction of the BDPA spins does produce a FID in either case, and the strong signal of BDPA makes this approach more convenient. The entire pulse sequence is $\mathrm{p}_\alpha$-$T$-$\mathrm{p}_{\pi/2}$-$\mbox{FID}$. The sequence is repeated 1024 times with a shot repetition time of \unit[1.02]{ms}. The duration of $\mathrm{p}_{\alpha}$ is then increased by \unit[4]{ns}, and 256 such steps are taken in order to vary $\mathrm{p}_{\alpha}$ over the range from \unit[2]{ns} to \unit[1024]{ns}. The entire experiment thus takes less than \unit[5]{min}, and the resulting signal-to-noise ratio is $>$50. In Fig.~\ref{fig2}(c) the integrated FID response is shown as a function of $t_\mathrm{P}$ for various values of applied MW power. The signal varies sinusoidally (with a superimposed dampened envelope that depends on the $B_1$ homogeneity \cite{anthony_cpw,sherwin_fel_bdpa}), with a period $\omega_1$ which we use to determine $B_1$. The envelope decay and its dependence on $B_1$ are used to estimate the $B_1$ homogeneity $\delta B_1\propto B_1$ (with a proportionality factor that depends on the resonator and the sample inside the resonator). The Fourier transforms of these time-domain transient nutation data, are shown in Fig.~\ref{fig2}(d), and exhibit a single frequency component ($f_1$). Fig.~\ref{fig2}(e) shows $B_1$ as a function of MW power. Fig.~\ref{fig2}(f) shows the dependence of the measured $B_1$ on the device orientation showing that device rotations can cause strong variations of $B_1$ between 0.3mT and 0.4mT, likely due to a small offset of the device position from the rotation axis within the resonator. This underlines the need for a measurement of $B_1$ for ISHE experiments as well as a procedure to determine the range of the $B_1$ distribution across the sample.

\begin{figure}
\includegraphics[width=\columnwidth]{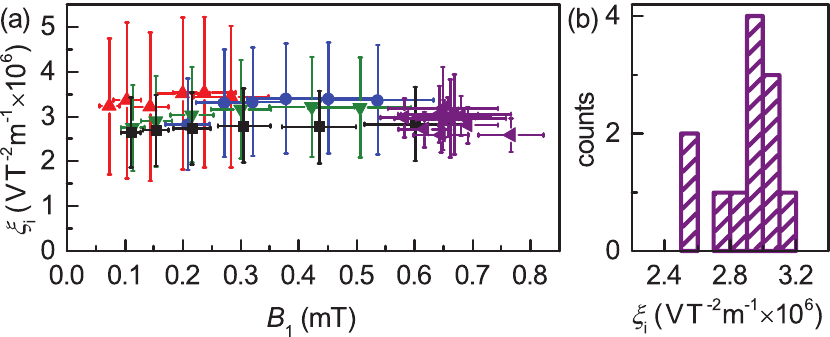}
\caption{\label{fig3} (a) Values of $\xi_\mathrm{i}$, determined from the measured $V_\mathrm{i}$ and $B_1$ for various experimental conditions as a function of $B_1$. The horizontal and vertical bars for each point do not represent statistical errors of $B_1$, but uncertainty intervals that are due to the distribution of $B_1$ throughout the sample. The points shown in purple were obtained under nominally identical conditions [cf.\ Fig.~\ref{fig1}(c)], whereas the other points were measured as a function of MW power at different resonator coupling adjustment settings (black: $Q\approx 200$, blue: $Q\approx 400$, green: $Q\approx 800$, red: $Q\approx 1000$) [cf.\ Fig.~\ref{fig1}(d)]. (b) Histogram of the individual $\xi_\mathrm{i}$ values for each repetition.}
\end{figure}

The procedure described above provides robust access to the absolute value of $B_1$ during ISHE experiments spectra even when the exact amount of MW power applied to the resonator, the resonator coupling, the $Q$ of the resonator, and the exact placement of device inside the resonator are not known. The strong EPR signal of BDPA enables fast acquisition of the transient nutation signal, and the integration of the BDPA into the ISHE device does not affect the ISHE measurements. We have routinely performed these measurements for all ISHE experiments, each time a device was installed, moved, rotated, or when the resonator coupling was altered. In Fig.~\ref{fig3}(a) the values of $\xi_\mathrm{i}=E_\mathrm{i}/B_1^2$ for the data shown in Fig.~\ref{fig1}(c,d) are shown as a function of $B_1$. The colors represent the various experimental conditions (in particular, the resonator coupling adjustment), and data points of the same color are measured under comparable conditions (apart from MW power). The horizontal bars represent $\delta B_1$, which is only limited by the inhomogeneity of the resonator. For the empty (unperturbed) resonator, $\delta B_1/B_1\approx 5\%$, whereas for resonator loaded with the ISHE template, we typically find $\delta B_1/B_1\approx 20\%$ showing that, in spite of a design of the ISHE device that aimed to keep $B_1$ homogeneous, ISHE samples introduce significant distortions of the resonator modes. The vertical bars are also governed by $\delta B_1$ since the error in $V_\mathrm{i}^\mathrm{c}$ is negligible. Within the confidence intervals, the values of $\xi_\mathrm{i}$ are constant, and are independent of the $B_1$ amplitude and the resonator coupling adjustment. Fig.~\ref{fig3}(b) shows a histogram of the measured $\xi_\mathrm{i}$ values obtained from the data shown in Fig.~\ref{fig1}(c). The values lie on a much more narrow distribution with a relative standard deviation of 6.6\%. This result shows that the application of the $B_1$ monitoring procedure leads to a considerably better reproducibility of quantitative ISHE measurements and, more importantly, an accurate absolute value is determined which characterizes the spin pumping and spin transport behavior of the given interface. Furthermore, the residual variation observed for the measured values of $\xi_\mathrm{i}$ can be attributed to the MW field inhomogeneities which, by optimization of the design of the sample and resonator geometries may be further improved.

In conclusion, it has been demonstrated how to accurately determine the magnetic  driving field strength $B_1$ of the FMR during an ISHE experiment by transient nutation measurement of a BDPA  standard integrated in the ISHE device. This measurement of $B_1$ is independent of the placement and orientation of the ISHE device in the MW field, the resonator coupling adjustment, and MW power and its accuracy relies solely on the homogeneity of $B_1$ throughout the ISHE device. We demonstrated the fidelity of this approach by statistical evaluation of a set of nominally identical measurements of the ISHE response of a device with subsequent measurement of $B_1$, and by deliberate variation of MW power and resonator coupling adjustment. It further shows that the ratio $\xi_\mathrm{i}=E_\mathrm{i}/B_1^2$ of the ISHE-electric field $E_\mathrm{i}$ to $B_1^2$ is characteristic for a given ISHE experiment. For quantitative studies of spin transport using FMR pumped spin current and ISHE detection, $\xi_\mathrm{i}$ appears to be a significantly better parameter for the characterization of a given ferromagnet to non-ferromagnet interface compared to just the ISHE voltage which is highly dependent on the experimental conditions, including device and resonator geometries, resonator tuning, and the applied MW driving field power. The accuracy of $\xi_\mathrm{i}$ is mainly limited by the inhomogeneity of $B_1$, which is a property of the waveguide or resonator that produces the MW field at the position of the sample, along with the sample structure that can distort these fields. This ultimately limits the quantitative interpretation of ISHE experiments, and great care must be taken to limit the inhomogeneity of the MW field.
\begin{acknowledgments}
This work was supported by the National Science Foundation (DMR-1404634). We also acknowledge the NSF Material Science and Engineering Center at the University of Utah (DMR-1121252) for supporting the device preparation facilities.
\end{acknowledgments}
\nocite{*}
\bibliography{refs}
\end{document}